\documentclass[12pt,a4paper]{article}
\usepackage{cmap}
\usepackage[cp1251]{inputenc}
\usepackage[english]{babel}
\usepackage{amsmath,amsthm,amssymb}
\usepackage{amsopn}
\usepackage{bm} 
\usepackage{cite}
\usepackage[margin=25mm]{geometry} 

\newtheorem{lemma}{Lemma}
\newtheorem{theorem}{Theorem}
\newtheorem{proposition}{Proposition}

\usepackage{mathtools}
\providecommand\given{} 
\newcommand\SetSymbol[1][]{%
	\nonscript\:#1\vert
	\allowbreak \nonscript\:	\mathopen{}}
\DeclarePairedDelimiterX\Set[1]\{\}{%
	\renewcommand\given{\SetSymbol[\delimsize]}	#1} 

\newcommand{\N}{\mathbb{N}}  

\newcommand{\BQP}{\textup{BQP}} 
\newcommand{\QAP}{\textup{QAP}} 
\newcommand{\TSP}{\textup{TSP}} 

\usepackage[colorlinks, linkcolor=blue, pagecolor=blue, urlcolor=black]{hyperref}

\begin{document}
\title{Affine maps between quadratic assignment polytopes\\ and subgraph isomorphism polytopes%
	\thanks{Supported by the~State Assignment for Research in P.G. Demidov Yaroslavl State University, 1.5768.2017/P220.}}
\author{A.N. Maksimenko\\[2ex]
\normalsize
P.G. Demidov Yaroslavl State University, 150000, Yaroslavl, Sovetskaya 14\\
\normalsize
\url{maximenko.a.n@gmail.com}}

\maketitle

\begin{abstract}
We consider two polytopes.
The quadratic assignment polytope $\QAP(n)$ is the convex hull of the set of tensors $x\otimes x$, $x \in P_n$, where $P_n$ is the set of $n\times n$ permutation matrices.
The second polytope is defined as follows.
For every permutation of vertices of the complete graph $K_n$ we consider appropriate $\binom{n}{2} \times \binom{n}{2}$ permutation matrix of the edges of $K_n$.
The Young polytope $P((n-2,2))$ is the convex hull of all such matrices.

In 2009, S. Onn showed that the subgraph isomorphism problem can be reduced to optimization both over $\QAP(n)$ and over $P((n-2,2))$.
He also posed the question whether $\QAP(n)$ and $P((n-2,2))$, having $n!$ vertices each, are isomorphic.
We show that $\QAP(n)$ and $P((n-2,2))$ are not isomorphic.
Also, we show that $\QAP(n)$ is a face of $P((2n-2,2))$,
but $P((n-2,2))$ is a projection of $\QAP(n)$.
\end{abstract}

\emph{The Boolean quadratic polytope} is the convex hull of the set
\[
\BQP(n) = \Set*{\bm{x}\otimes\bm{x} \given \bm{x} \in \{0,1\}^n},
\]
where $\bm{x}\otimes\bm{x} \in \{0,1\}^{n\times n}$ is the tensor product and $n\in\N$, $n \ge 2$.
All polytopes considered in this paper are defined as convex hulls of sets of vertices.
Therefore, we introduce the notation only for the set of vertices, and not for the whole polytope, and we often call the set of vertices a polytope.

Recently, the properties of Boolean quadratic polytopes and of the affinely equivalent to them cut polytopes are studied quite intensively.
In particular, this is confirmed by the large number of citations\footnote{The search engine scholar.google.com reports 895 citations on May 29, 2017} of the  Deza and Laurent monograph~\cite{Deza:1997}, entirely devoted to this topic.
It is known that $\BQP(n)$ is 3-neighborly~\cite{Deza:1992} (every three vertices form a face of the polytope), it has an exponential complexity of an extended formulation~\cite{Fiorini:2015, Kaibel:2015}, it is affinely equivalent to some faces of polytopes associated with many other NP-hard problems (the traveling salesman problem, the knapsack problem, the set covering and set packing problems, the maximal 3-satisfiability problem, the 3-assignment problem, and many others)~\cite{Maksimenko:2016}.
It is also known that the polytopes of any linear combinatorial optimization problem are affine projections of faces of Boolean quadratic polytopes~\cite{Maksimenko:2012}.

Following~\cite{Onn:2009}, we denote the set of permutation $n\times n$-matrices by ${\cal P}_n$.
\emph{The quadratic assignment polytope} is the convex hull of the set
\[
\QAP(n) = \Set*{\bm{y}\otimes\bm{y} \given \bm{y}\in{\cal P}_n}.
\]
In~\cite{Onn:2009}, the notation $\psi_n$ is used for the convex hull of $\QAP(n)$.
There is also described that deciding if a graph $G$ has a subgraph isomorphic to a graph $H$ is reduced to optimization over $\QAP(n)$.

The following theorem has a simple proof (see the next page) and is mentioned in several works.

\begin{theorem}[\cite{Kaibel:1997,Rijal:1995,Saito:2009}]
	\label{thm:1}
	The polytope $\QAP(n)$ is affinely isomorphic to a face of $\BQP(n^2)$.
\end{theorem}

The subgraph isomorphism problem can also be reduced to optimization over the Young polytope $P((n-2,2))$ defined as follows~\cite{Onn:2009}.
For each permutation of the vertices of the complete graph $K_n$, consider the corresponding $\binom{n}{2}\times\binom{n}{2}$-matrix of permutations of edges of $K_n$.
We denote by $\phi_n$ the set of all such matrices for a fixed  $n\in\N$, $n\ge 3$.
The convex hull of $\phi_n$ is a particular example of a Young polytope and is denoted by $P((n-2,2))$ in~\cite{Onn:1993}. It is a 2-neighborly polytope~\cite{Onn:1993}.
It is easy to show that the polytope $\phi_n$ is a projection of $\QAP(n)$ (Proposition~\ref{prop:1} below).

In~\cite{Onn:2009}, S. Onn posed the question
whether $\phi_n$ and $\QAP(n)$, having $n!$ vertices each, are isomorphic.

We show that $\phi_n$ is not 3-neighborly.
For $n = 3$, it is verified directly.
For $n > 3$, we use the following

\begin{lemma}
	\label{lem:1}
	The polytope $\phi_3$ is affinely isomorphic to a face of $\phi_n$.
\end{lemma}

On the other hand, from Theorem~\ref{thm:1} and the 3-neighborliness of $\BQP(n)$ it follows that $\QAP(n)$ is also 3-neighborly.
Hence, $\phi_n$ and $\QAP(n)$ are not isomorphic.

The polytopes $\phi_n$ and $\QAP(n)$, in particular, are interesting because the traveling salesman polytope $\TSP(n)$ is a projection of each of them.
Taking into account that the Boolean quadratic polytope $\BQP(k)$ is a face of $\TSP(n)$ for $n = 2k (2k-1)$~\cite{Maksimenko:2013}, one can conclude that $\BQP(k)$ is a projection of some face of the polytope $\phi_n$ (the Young polytope~$P((n-2,2))$).
It turns out that there is a closer connection between $\BQP(k)$ and $\phi_n$.

\begin{theorem}
	\label{thm:2}
	The polytope $\BQP(k)$ is affinely isomorphic to a face of $\phi_{2k}$.
\end{theorem}

In particular, $\phi_n$ has a $3$-neighborly face with $2^{\lfloor n/2\rfloor}$ vertices.
Moreover, according to~\cite{Maksimenko:2014}, for any $s\in\N$ and $k \ge 2^{\lceil 2s/3\rceil}$ the polytope $\BQP(k)$ and with it the polytope $\phi_{2k}$
have an $s$-neighborly face with superpolynomial (w.r.t $k$) number of vertices.

\section{Proofs}

For the coordinates of the vector $\bm{x} \in \BQP(m) \subset \{0,1\}^{m^2}$ we introduce the notation $x_{ij}$, $i,j\in[m]$. Therefore,
\[
\BQP(m) = \Set*{\bm{x} \in \{0,1\}^{m^2}\given x_{ij} = x_{ii} x_{jj},\ i,j\in[m]}.
\]

A vector from the set $\QAP(n) \subset \{0,1\}^{n^4}$ is denoted by $\bm{y}$, its coordinates are denoted by $y_{ijkl}$, $i,j,k,l \in [n]$. 
By the definition, $\QAP(n)$ consists of the vectors $\bm{y} \in \{0,1\}^{n^4}$, satisfying the equations
\begin{align}
&y_{ijkl} = y_{ijij} y_{klkl}, && i,j,k,l \in [n], \label{eq:psi1}\\
&\sum_{j\in[n]} y_{ijij} = 1,  && i \in [n], \label{eq:psi2}\\
&\sum_{i\in[n]} y_{ijij} = 1,  && j \in [n]. \label{eq:psi3}
\end{align}

\begin{proof}[\normalfont\textbf{Proof of Theorem~\ref{thm:1}}]
For the coordinates of the vector $\bm{x} \in \BQP(n^2)$ we use the notation $x_{ijkl}$, $i,j,k,l \in [n]$.
According to the definition of $\BQP(m)$,
\begin{equation}
\label{eq:BQPpsi1}
x_{ijkl} = x_{ijij} x_{klkl}, \qquad \bm{x} \in \BQP(n^2).
\end{equation}

Consider the face $F_1$ of $\BQP(n^2)$ lying in the intersection of the supporting hyperplanes
\begin{equation}
\label{eq:psi2BQP}
x_{ijil} = 0,  \qquad  i,j,l \in [n], \quad j \neq l.
\end{equation}
Since $x_{ijil} = x_{ijij} x_{ilil}$ for all $\bm{x} \in \BQP(n^2)$, then the restrictions~\eqref{eq:psi2BQP} are equivalent to the inequalities
\[
\sum_{j\in[n]} x_{ijij} \le 1, \qquad i \in [n].
\]
Setting
\begin{equation}
\label{eq:BQPpsi2}
\sum_{j\in[n]} x_{ijij} = 1, \qquad i \in [n],
\end{equation}
we proceed to consider the face $F_2$, $F_2 \subset F_1$.

Analogically, the equations
\begin{equation}
\label{eq:BQPpsi3}
\sum_{i\in[n]} x_{ijij} = 1, \qquad j \in [n],
\end{equation}
define some face $F_3$, $F_3 \subset F_2$, of the polytope $\BQP(n^2)$.

Comparing the equations \eqref{eq:BQPpsi1}, \eqref{eq:BQPpsi2}, \eqref{eq:BQPpsi3} and the equations \eqref{eq:psi1}--\eqref{eq:psi3}, we come to the conclusion, that the face $F_3$ is affinely isomorphic to $\QAP(n)$.
\end{proof}


Let us now consider the polytope $\phi_n$.
Coordinates of a vector $\bm{z} \in \phi_n \subset \{0,1\}^{\binom{n}{2} \times \binom{n}{2}}$ are denoted by $z_{ijkl}$, $i,j,k,l \in [n]$, $i < j$, $k < l$.
In particular, 
\begin{align*}
\sum_{1 \le k < l \le n} z_{ijkl} &= 1, \qquad 1 \le i < j \le n,\\
\sum_{1 \le i < j \le n} z_{ijkl} &= 1, \qquad 1 \le k < l \le n,
\end{align*}
for $\bm{z} \in \phi_n$.

The validity of the following statement is obvious.

\begin{proposition}
	\label{prop:1}
	The polytope $\QAP(n)$ is projected onto $\phi_n$ by the affine map
	\(y_{ikjl}+y_{iljk} \mapsto z_{ijkl}\), $i,j,k,l\in[n]$, $i < j$, $k < l$.
\end{proposition}

Let us show, that $\phi_n$ is not 3-neighborly.

The set $\phi_3 \subset \{0,1\}^{3\times3}$ consists of six vectors
\[
\begin{pmatrix}
1 0 0\\
0 1 0\\
0 0 1\\
\end{pmatrix}, 
\quad
\begin{pmatrix}
0 0 1\\
1 0 0\\
0 1 0\\
\end{pmatrix},
\quad
\begin{pmatrix}
0 1 0\\
0 0 1\\
1 0 0\\
\end{pmatrix},
\quad
\begin{pmatrix}
0 0 1\\
0 1 0\\
1 0 0\\
\end{pmatrix},
\quad
\begin{pmatrix}
1 0 0\\
0 0 1\\
0 1 0\\
\end{pmatrix},
\quad
\begin{pmatrix}
0 1 0\\
1 0 0\\
0 0 1\\
\end{pmatrix}.
\]
It is easy to verify that the sum of the first three vectors is equal to the sum of the last three.
This means that the first three vectors (and also the last three) do not form a face of $\phi_3$.

\begin{proof}[\normalfont\textbf{Proof of Lemma~\ref{lem:1}}]
We consider the face $F$ of the polytope $\phi_n$ lying in the intersection of the supporting hyperplanes
\[
z_{ijkl} = 0, \qquad j > 3, \quad l \neq j, \quad 1 \le i < j \le n, \quad 1 \le k < l \le n.
\]
In particular, $z_{ijij} = 1$ for $3 < i < j \le n$ and $\bm{z} \in F$.
That is, we consider only permutations of the vertices of the corresponding graph $K_n$ that do not change the positions of the vertices 4, 5, \dots, $n$.

Thus, for every $\bm{z}\in F$ the following is valid:
\[
(i > 3 \text{ and } k \neq i) \Rightarrow
(z_{ijkl} = 0), \qquad \forall i,j,k,l \in [n],\ i < j,\ k < l.
\]
Similar statements are valid $k > 3$ and $i \neq k$, 
and also for $l > 3$ and $j \neq l$.

Therefore, the coordinates $z_{ijkl}$ are not fixed, if they satisfy the following conditions: \
1) $i,k \in [3]$; \ 2) $j,l \in [3]$ or $j = l$.

To complete the proof, it remains to show that for $\bm{z} \in F$ the coordiantes $z_{ijkj}$, $i,k \in [3]$, $j \in \{4,5,\dots,n\}$,
are expressed linearly in terms of $z_{ijkl}$, $i,j,k,l \in [3]$.

We introduce the auxiliary variable
\[
h_{12} = z_{1212} + z_{1223} + z_{1312} + z_{1323} - 1.
\]
Note that for all $\bm{z}\in F$ the following holds:
\[
0 \le h_{12} \le 1.
\]
Moreover, $h_ {12} = 1$ if and only if the permutation of the vertices of the corresponding graph $K_n$ performs the replacement $1 \to 2$.
Hence, $z_{1j2j} = h_{12}$ for $j \in \{4,5,\dots,n\}$.
In a similar way,
variables $h_{ik}$, $i,k \in [3]$, are defined
and $z_{ijkj} = h_{ik}$, $i,k \in [3]$, $j \in \{4,5,\dots,n\}$.
\end{proof}

\begin{proof}[\normalfont\textbf{Proof of Theorem~\ref{thm:2}}]
We consider the face $F$ of the polytope $\phi_{2n}$ lying in the intersection of the supporting hyperplanes
\[
z_{2i-1, 2i, 2i-1, 2i} = 1, \qquad i\in[n].
\]
Obviously, in this way we fix the following restriction on the vectors from $F$.
The corresponding permutation of the vertices of the graph $K_n$ either reverses the vertices $2i-1$ and $2i$, or leaves them unchanged.

It is easy to verify that the coordinates of the vector $\bm{z} \in F$ satisfy the following conditions:
\begin{gather*}
z_{1,3,1,3} = 
z_{1,4,1,4} = 
z_{2,3,2,3} = 
z_{2,4,2,4},\\
z_{1,3,1,4} = 
z_{1,4,1,3} = 
z_{2,3,2,4} = 
z_{2,4,2,3},\\
z_{1,3,2,3} = 
z_{1,4,2,4} = 
z_{2,3,1,3} = 
z_{2,4,1,4},\\
z_{1,3,2,4} = 
z_{1,4,2,3} = 
z_{2,3,1,4} = 
z_{2,4,1,3},\\
z_{2,4,1,3} +
z_{2,4,1,4} +
z_{2,4,2,3} +
z_{2,4,2,4} = 1.
\end{gather*}
Obviously, all these equations remain valid if we simultaneously replace index 1 by $2i-1$, 2 by $2i$, 3 by $2j-1$, 4 by $2j$, where $i,j \in [n]$, $i < j$.

Hence, all the coordinates of $\bm{z} \in F$ are expressed linearly in terms of 
$3 \cdot \binom{n}{2}$ coordinates $z_{2i,2j,2i-1,2j}$, $z_{2i,2j,2i,2j-1}$, and $z_{2i,2j,2i,2j}$, $i,j \in [n]$, $i < j$.

Note, that the value of the sum $z_{2i,2j,2i,2j} + z_{2i,2j,2i,2j-1}$ is an indicator of the fact that the appropriate permutation of the vertices of $K_n$ leaves $2i-1$ and $2i$ unchanged.
Similar, $z_{2i,2j,2i,2j} + z_{2i,2j,2i-1,2j}$ is equal to 1 if and only if the vertices $2j-1$ and $2j$ are not permuted. 
In other words, for any $i, j \in [n]$, $i < j$, we have 
\[
z_{2i,2j,2i,2j} = 1 \iff
z_{2i,2j,2i,2j} + z_{2i,2j,2i,2j-1} = z_{2i,2j,2i,2j} + z_{2i,2j,2i-1,2j} = 1
\]
and
\begin{align*}
z_{2i,2j,2i,2j} + z_{2i,2j,2i,2j-1} &= z_{2i,2l,2i,2l} + z_{2i,2l,2i,2l-1}, 
&& \forall l \in [n], \ i<l,\\
z_{2i,2j,2i,2j} + z_{2i,2j,2i-1,2j} &= z_{2k,2j,2k,2j} + z_{2k,2j,2k-1,2j},
&& \forall k \in [n], \ k<j.
\end{align*}

Therefore, the polytope $\BQP(n)$ and the described face $F$ of $\phi_{2n}$ are connected by the equations
\begin{align*}
z_{2i,2j,2i,2j} &= x_{ij}, \\  
z_{2i,2j,2i,2j-1} &= x_{ii} - x_{ij}, \\
z_{2i,2j,2i-1,2j} &= x_{jj} - x_{ij}, \\
\end{align*}
where $i,j\in[n]$, $i < j$.
\end{proof}

\end{document}